\documentclass[a4paper,12pt]{article}
\usepackage{jheppub2} % for details on the use of the package, please
\usepackage[T1]{fontenc} % if needed

\author[a]{Marco Bochicchio}
\affiliation[a]{INFN sez. Roma 1\\Piazzale A. Moro 2, Roma, I-00185, Italy}

\emailAdd{marco.bochicchio@roma1.infn.it}

\abstract{ YM and QCD are known to be renormalizable, but not ultraviolet finite, order by order in perturbation theory. It is a fundamental question as to whether YM or QCD are ultraviolet finite, or only renormalizable,
order by order in the large-N 't Hooft or Veneziano expansions. We demonstrate that Renormalization Group and Asymptotic Freedom imply that in 't Hooft large-N expansion the S-matrix in YM is ultraviolet finite, while in both  't Hooft and Veneziano large-N expansions the S-matrix in confining QCD with massless quarks is renormalizable but not ultraviolet finite. By the same argument it follows that the large-N $\mathcal{N}=1$ SUSY YM S-matrix is ultraviolet finite as well. Besides, we demonstrate that the correlators of local gauge-invariant operators, as opposed to the S-matrix, are renormalizable but in general not ultraviolet finite in the large-N 't Hooft and Veneziano expansions, neither in pure YM and $\mathcal{N}=1$ SUSY YM nor a fortiori in massless QCD. Moreover, we compute explicitly the counterterms that arise renormalizing the large-N 't Hooft and Veneziano expansions, by deriving in confining massless QCD-like theories a low-energy theorem of NSVZ type, that relates the log derivative with respect to the gauge coupling of a $k$-point correlator, or the log derivative with respect to the RG-invariant scale, to a $k+1$-point correlator with the insertion of $\Tr F^2$ at zero momentum. Finally, we argue that similar results hold in the large-N limit of a vast class of confining QCD-like theories with massive matter fields, provided a renormalization scheme exists, as for example $\overline{MS}$, in which the beta function is independent on the masses. In particular, in both 't Hooft and Veneziano large-N expansions the S-matrix in confining massive QCD and massive $\mathcal{N}=1$ SUSY QCD is renormalizable but not ultraviolet finite. }

\usepackage{braket}
\usepackage{amsmath}
\usepackage{amssymb}
\usepackage{graphicx}
\usepackage{hyperref}
\usepackage{fancyhdr}
\def\beq{\begin{equation}}
\def\eeq{\end{equation}}
\def\bea{\begin{eqnarray}}
\def\eea{\end{eqnarray}}
\def\bq{\begin{quote}}
\def\eq{\end{quote}}
\DeclareMathOperator{\Tr}{Tr}

\title{The large-N Yang-Mills S-matrix is ultraviolet finite, but the large-N QCD S-matrix is only renormalizable}
\date{}
\begin{document}
\maketitle
\section{Introduction}

SU(N) Yang-Mills (YM) and SU(N) QCD with $N_f$ quark flavors are known to be renormalizable but not ultraviolet finite in perturbation theory. It is a fundamental question, that has never been considered previously, as to whether their large-N 't Hooft or Veneziano expansions (Section \ref{1}) enjoy better ultraviolet properties non-perturbatively, perhaps limiting only to the large-N S-matrix, once the lowest $\frac{1}{N}$ order has been made finite by renormalization as defined in Sections \ref{2}, \ref{3}. Answering this question sets the strongest constraints on the solution, that is yet to come, of large-N YM and QCD. \par
The first main result in this paper is that Renormalization Group (RG) and Asymptotic Freedom (AF) imply that in 't Hooft expansion the large-N YM S-matrix is ultraviolet finite, while in both 't Hooft and Veneziano expansions the large-N S-matrix in confining massless QCD \footnote{By massless QCD we mean QCD with massless quarks.} is renormalizable but not ultraviolet finite (Section \ref{2}): In 't Hooft expansion due to log divergences of meson loops (Section \ref{1}) starting at order of $\frac{N_f}{N}$, in Veneziano expansion due to loglog divergences of "overlapping" meson-glueball loops (Section \ref{1}) starting at order of $\frac{N_f}{N^3}$. By the same argument it follows that in 't Hooft expansion the large-N $\mathcal{N}=1$ SUSY YM S-matrix is ultraviolet finite as well.\par
Correlators (Section \ref{3}), as opposed to the S-matrix, turn out to be renormalizable but loglog divergent in general, in addition to the possible divergences of the S-matrix in the aforementioned large-N expansions, but at the lowest order, even in pure large-N YM and $\mathcal{N}=1$ SUSY YM. \par
The second main result is a low-energy theorem (Section \ref{4}) of Novikov-Shifman-Vainshtein-Zakharov (NSVZ) type in confining massless QCD-like theories \footnote{By QCD-like theory we mean a confining Asymptotically Free (AF) gauge theory admitting the large-N 't Hooft or Veneziano limits. We call such a theory massive if its matter fields are massive, and massless if a choice of parameters exists for which the theory is massless to all orders perturbation theory.}, that allows us to compute explicitly the lowest-order large-N counterterms implied by RG and AF as opposed to perturbation theory. \par
Finally, we argue that similar results hold (Section \ref{5}) for the large-N S-matrix in a vast class of confining QCD-like theories with massive matter fields, provided a renormalization scheme exists in which the beta function is independent on the masses. $\overline{MS}$ is an example of such a scheme. Besides, the asymptotic results in Section \ref{3} extend also to the correlators of the massive theory provided the massless limit of the massive theory exists smoothly.

\section{Large-N 't Hooft and Veneziano expansions} \label{1}
We recall briefly the 't Hooft \cite{H1} and Veneziano \cite{Veneziano0} expansions in large-N YM and QCD with $N_f$ quark flavors. \par
Non-perturbatively, 't Hooft large-N limit is defined computing the QCD functional integral in a neighborhood of $N=\infty$ with 't Hooft gauge coupling $g^2 = g^2_{YM} N$ and $N_f$ fixed. The corresponding perturbative expansion, once expressed in terms of $g^2$, can be reorganized in such a way that each power of $\frac{1}{N}$ contains the contribution of an infinite series in $g^2$ \cite{H1,Veneziano0}. \par
The lowest-order contribution in powers of $\frac{1}{N}$ to connected correlators of local single-trace gauge-invariant operators $\mathcal{G}_i(x_i)$ and of quark bilinears $\mathcal{M}_i(x_i)$, both normalized in such a way that the two-point correlators are on the order of $1$, turns out to be on the order of:
\bea \label{P}
&&\langle \mathcal{G}_1(x_1)\mathcal{G}_2(x_2)\cdots \mathcal{G}_n(x_n) \rangle_{conn}\sim N^{2-n} \, ; \, 
\langle\mathcal{M}_1(x_1)\mathcal{M}_2(x_2)\cdots \mathcal{M}_k(x_k)\rangle_{conn}\sim N^{1-\frac{k}{2}} \nonumber \\
&&\langle \mathcal{G}_1(x_1)\mathcal{G}_2(x_2)\cdots \mathcal{G}_n(x_n)\mathcal{M}_1(x_1)\mathcal{M}_2(x_2)\cdots \mathcal{M}_k(x_k)\rangle_{conn}\sim N^{1-n-\frac{k}{2}} 
\eea
This is the 't Hooft Planar Theory, that perturbatively sums Feynman graphs triangulating respectively a sphere with $n$ punctures, a disk with $k$ punctures on the boundary, and a disk with $k$ punctures on the boundary and $n$ punctures in the interior. The punctured disk arises in 't Hooft large-N expansion from Feynman diagrams whose boundary is exactly one quark loop. \par
Higher-order contributions correspond to summing the Feynman graphs triangulating orientable Riemann surfaces with smaller fixed Euler characteristic. They correct additively 't Hooft Planar Theory with a weight $N^{\chi}$, where $\chi=2-2g-h-n-\frac{k}{2}$ is the Euler characteristic of an orientable Riemann surface of genus $g$ (i.e. a sphere with $g$ handles), with $h$ holes (or boundaries), $n$ marked points in the interior, and $k$ marked points on the boundary of some hole, that the Feynman graphs triangulate. Non-perturbatively a handle is interpreted as a glueball loop, and a hole as a meson loop \cite{H1,Veneziano0}.\par
On the contrary, non-perturbatively Veneziano large-N limit is defined computing the QCD functional integral in a neighborhood of $N=\infty$ with $g^2$ and $\frac{N_f}{N}$ fixed. Since in large-N QCD factors of the ratio $\frac{N_f}{N}$, that is kept fixed, may arise perturbatively only from quark loops, Veneziano large-N expansion contains perturbatively already at the lowest order Feynman graphs that triangulate a punctured sphere or a punctured disk with any number of holes, i.e. it contains the sum of all the Riemann surfaces that are geometrically planar: This is the Veneziano Planar Theory. Higher orders contain higher-genus Riemann surfaces.

\section{Large-N YM and massless QCD S-matrix} \label{2}
We assume that YM and QCD have been regularized in a way that we leave undefined, but in special cases in Section \ref{5}, by introducing a common cutoff scale $\Lambda$, perturbatively, in the large-N expansion, and non-perturbatively. The details of the regularization do not matter for our arguments. \par
In perturbation theory, pure YM and massless QCD need only gauge-coupling renormalization in the classical action in order to get a finite large-$\Lambda$ limit, since in massless QCD there is no quark-mass renormalization because chiral symmetry is exact in perturbation theory. In addition, local gauge invariant operators need also in general multiplicative renormalizations, associated to the anomalous dimensions of the operators, in order to make their correlators finite. \par
We will see in Section \ref{3} that also in large-N YM and massless QCD non-Planar multiplicative renormalizations occur in general in both 't Hooft and Veneziano expansions, once the Planar correlators (i.e. the lowest-order correlators) have been made finite by the Planar gauge-coupling and multiplicative renormalizations. \par
However, multiplicative renormalizations must cancel in the S-matrix because of the LSZ reduction formulae, since the S-matrix cannot depend on the choice of the interpolating fields for a given asymptotic state in the external lines \cite{AFB} (see also Section \ref{3}). Therefore, only gauge-coupling renormalization is necessary in the large-N YM and massless QCD S-matrix, but non-perturbatively according to the RG \footnote{We assume that the aforementioned theories actually exist mathematically and are renormalizable, that the $\frac{1}{N}$ expansion is at least asymptotic, and that standard RG is actually asymptotic in the ultraviolet to the exact result because of asymptotic freedom. Though these statements are universally believed, no rigorous mathematical construction of YM or of QCD or of their large-N limits presently exists, let alone a mathematically rigorous proof of these statements.}, because of the summation of an infinite number of Feynman graphs at any fixed $\frac{1}{N}$ order. 
Non-perturbatively, gauge-coupling renormalization is equivalent to make finite and (asymptotically) constant the RG-invariant scale: $\Lambda_{RG}= const \Lambda \exp(-\frac{1}{2\beta_0 g^2}) (\beta_0 g^2)^{-\frac{\beta_1}{2 \beta_0^2}}(1+...)$, uniformly for arbitrarily large $\Lambda$ in a neighborhood of $g=0$, where the dots represent 
an asymptotic series in $g^2$ of renormalization-scheme dependent terms, that obviously vanish as $g \rightarrow 0$. The overall constant is scheme dependent as well. \par 
Moreover, non-perturbatively RG requires that every physical mass scale of the theory is proportional to $\Lambda_{RG}$. Therefore, being $\Lambda_{RG}$ the only parameter occurring in the $S$-matrix in both large-N YM and confining massless QCD, the ultraviolet finiteness of the large-N S-matrix is equivalent to the existence of a renormalization scheme for $g$ in which the large-N expansion of $\Lambda_{RG}$ is finite. This is decided as follows.\par
We consider first 't Hooft expansion in large-N YM. In this case, $\beta_0= \beta_0^P= \frac{1}{(4 \pi)^2} \frac{11}{3}$, $\beta_1= \beta_1^P= \frac{1}{(4 \pi)^4} \frac{34}{3}$, where the superscript $P$ stands for 't Hooft Planar. Now, both in the 't Hoof Planar Theory and to all the $\frac{1}{N}$ orders, the first-two coefficients of the beta function $\beta_0, \beta_1$ get contributions only from 't Hooft Planar diagrams. This implies that in large-N YM the $\frac{1}{N}$ expansion of $\Lambda_{YM}$ is in fact finite \cite{AF}, the non-Planar $\frac{1}{N}$ corrections occurring in the dots or in $const$ contributing only at most a finite change of renormalization scheme to the 't Hooft Planar RG-invariant  scale, $ \Lambda^P_{YM} =const  \Lambda \exp(-\frac{1}{2\beta^{P}_0 g^2}) (\beta^{P}_0 g^2)^{-\frac{\beta^{P}_1}{2 \beta_0^{P2}}}(1+...)$. \par
Thus the S-matrix in large-N YM is finite in 't Hooft expansion around the Planar Theory, once the Planar Theory has been made finite by the gauge-coupling renormalization implicit in the finiteness of $\Lambda^P_{YM}$ \cite{AF}. Indeed, since YM is renormalizable, all glueball loops must be finite in the S-matrix (i.e. on-shell, see also Section \ref{3}), because if they were divergent, their divergence ought to be reabsorbed into a divergent redefinition of $\Lambda_{YM}$, that is the only parameter in the $S$-matrix, contrary to what we have just shown. A similar argument implies that in 't Hooft expansion the large-N $\mathcal{N}=1$ SUSY YM S-matrix is ultraviolet finite as well.\par
't Hooft expansion of large-N massless QCD is deeply different.
In this case, $\beta_0= \beta_0^P + \beta_0^{NP}= \frac{1}{(4 \pi)^2} \frac{11}{3} - \frac{1}{(4 \pi)^2} \frac{2}{3} \frac{N_f}{N}$ and $\beta_1= \beta_1^P + \beta_1^{NP}= \frac{1}{(4 \pi)^4} \frac{34}{3} - \frac{1}{(4 \pi)^4} (\frac{13}{3} - \frac{1}{N^2}) \frac{N_f}{N}$, where the superscript $NP$ stand for non-'t Hooft Planar.
Since quark loops occur at order of $\frac{1}{N}$, the first coefficient of the beta function, $\beta^P_0$, gets an additive non-'t Hooft Planar $\frac{1}{N}$ correction, $\beta_0^{NP}= - \frac{1}{(4 \pi)^2} \frac{2}{3} \frac{N_f}{N}$. As a consequence it is impossible to find a renormalization scheme for $g$ that makes $\Lambda_{QCD}$ finite in the 't Hooft Planar Theory and in the next order of the $\frac{1}{N}$ expansion at the same time, as the following computation shows \cite{AF}: 
\bea \label{alpha1}
&&\Lambda_{QCD} \sim  \Lambda \exp(-\frac{1}{2\beta^{P}_0 (1+\frac{\beta_0^{NP}}{\beta_0^{P}}) g^2 }) 
\sim  \Lambda \exp(-\frac{1}{2\beta^{P}_0  g^2 })(1+\frac{\frac{\beta_0^{NP}}{\beta_0^{P}}}{2\beta^{P}_0  g^2 }) \nonumber \\
&&\sim \Lambda^P_{QCD} (1+\frac{\beta_0^{NP}}{\beta_0^{P}} \log(\frac{\Lambda}{ \Lambda^P_{QCD}})) \eea
where in the first line $g$ is a bare free parameter according to the RG to all the $\frac{1}{N}$ orders, while in the last line we have renormalized $g$ according to the Asymptotic Freedom of the 't Hooft Planar Theory $\frac{1}{2\beta^{P}_0  g^2} \sim \log(\frac{\Lambda}{\Lambda^P_{QCD}})$, as follows for consistency by requiring that $\Lambda^P_{QCD}$ is finite uniformly in a neighborhood of $\Lambda =\infty$. \par
The symbol $\sim$ in this paper means asymptotic equality in a sense specified by the context, up to perhaps a non-zero constant overall factor. We should notice that the equalities in Equation \ref{alpha1} hold asymptotically, uniformly for large finite $\Lambda$ and small $g$ even before Planar renormalization, without the need to actually take the limits $\Lambda  \rightarrow \infty$, $g \rightarrow 0$, as they are obtained expressing $g$ identically in terms of $\Lambda^P_{QCD}$ in the last asymptotic equality. We emphasize that the log divergence in Equation \ref{alpha1} occurs precisely because of the Asymptotic Freedom of the Planar Theory. \par
In Section \ref{4} we will compute explicitly by means of a low-energy theorem the large-N counterterm due to the renormalization of $\Lambda^P_{QCD}$, that turns out to agree exactly, within the leading-log accuracy, with the perturbative counterterm due to quark loops. Indeed, were $\Lambda^P_{QCD}$ to get only a finite renormalization, the complete large-N QCD and the 't Hooft Planar Theory would have the same $\beta_0$, that is false. \par
Hence, being $\Lambda_{QCD}$ the only physical mass scale, glueball and meson masses receive $\frac{1}{N}$ log-divergent self-energy corrections proportional to the one of $\Lambda_{QCD}$, that can arise only from a log divergence of meson loops. This is a physical fact, that characterizes the meson interactions in the ultraviolet (UV), reflecting the corresponding perturbative quark interactions in the UV. 
Therefore, 't Hooft expansion of the QCD S-matrix, though renormalizable, starting at order of $\frac{N_f}{N}$ is log divergent, due to log divergences of meson loops. \par
The chances of finiteness would seem more promising in the Veneziano expansion. In this case, $\beta_0= \beta_0^{VP}= \frac{1}{(4 \pi)^2} \frac{11}{3} - \frac{1}{(4 \pi)^2} \frac{2}{3} \frac{N_f}{N}$ and $\beta_1= \beta_1^{VP} + \beta_1^{NVP}$, with $\beta_1^{VP}= \frac{1}{(4 \pi)^4}( \frac{34}{3} - \frac{13}{3}\frac{N_f}{N})$ and $\beta_1^{NVP}= \frac{1}{(4 \pi)^4}\frac{N_f}{N^3}$, where the superscripts $VP$ and $NVP$ stand for Veneziano Planar and non-Veneziano Planar.
Since the Veneziano Planar Theory contains already all quark loops, the first coefficient of the Veneziano Planar beta function
and of the complete beta function coincide. As a consequence there is no log divergence in the expansion of $\Lambda_{QCD}$. \par
Nevertheless, also in the Veneziano expansion it is impossible to find a renormalization scheme for $g$ in which both $\Lambda^{VP}_{QCD}$ and its $\frac{1}{N}$ corrections are finite at the same time, because of a loglog divergence starting at order of $\frac{N_f}{N^3}$ due to "overlapping"
glueball-meson loops, as the following computation shows:
\bea \label{Ven}
&&\Lambda_{QCD} \sim   \Lambda \exp(-\frac{1}{2\beta_0 g^2}) (g^2)^{-\frac{\beta^{VP}_1}{2 \beta_0^2}}  (g^2)^{-\frac{\beta^{NVP}_1}{2 \beta_0^2}} \nonumber \\
&&\sim   \Lambda \exp(-\frac{1}{2\beta_0 g^2}) (g^2)^{-\frac{\beta^{VP}_1}{2 \beta_0^2}}  (1- \frac{\beta^{NVP}_1}{ 2\beta_0^2} \log g^2) \nonumber \\
&&\sim \Lambda^{VP}_{QCD} (1+\frac{\beta^{NVP}_1}{ 2\beta_0^2} \log\log(\frac{\Lambda}{ \Lambda^{VP}_{QCD}}))
\eea
Thus the large-N Veneziano expansion of the S-matrix in confining \footnote{In fact, Equation \ref{Ven} may be valid only for $\frac{N_f}{N}$ and $g$ in a certain neighborhood of $0$. Indeed, it is believed that there is a critical value of $\frac{N_f}{N}$ and of $g$ at which massless QCD becomes exactly conformal because of an infrared zero of the beta function. At this critical value of $\frac{N_f}{N}$ and of $g$, $\Lambda_{QCD}$ may vanish due to the infrared zero. Similar considerations may apply to other massless QCD-like theories (Section \ref{5}).} massless QCD is not ultraviolet finite as well. In any case both 't Hooft and Veneziano expansions of the S-matrix are renormalizable, all the aforementioned divergences being reabsorbed order by order in the $\frac{1}{N}$ expansions by a redefinition of $\Lambda_{QCD}$.

\section{Large-N YM and massless QCD correlators} \label{3}

We study now the multiplicative renormalizations of gauge-invariant operators in the large-N 't Hooft and Veneziano expansions. They are sufficient to make the correlators finite, once the gauge coupling and $\Lambda_{QCD}$ have been renormalized as described in Section \ref{2}, in any massless QCD-like theory. \par The computations greatly simplify if we reconstruct the asymptotic structure of the bare correlators from the asymptotic renormalized correlators, either in the complete theory or in the large-N expansions. In order to do so, we employ an asymptotic structure theorem \cite{MBN} for glueball and meson two-point correlators in 't Hooft large-N limit of massless QCD-like theories,
and the associated, but much more general, asymptotic estimates \cite{MBN,MBM}, that hold both in the complete theory and a fortiori in 't Hooft and Veneziano large-N limits.\par
For the aims of this paper it is sufficient to report the asymptotic theorem  \cite{MBN} in the coordinate representation. Under mild assumptions, it reads as follows.  
The connected two-point Euclidean correlator of a hermitian local single-trace gauge-invariant operator or of a quark bilinear, $\mathcal{O}^{(s)}$, of spin $s$, naive mass dimension $D$, and with anomalous dimension $\gamma_{\mathcal{O}^{(s)}}(g)$ \footnote{We suppose that the matrix of anomalous dimensions has been diagonalized, as generically possible at least at the leading order, which is the only one that matters for the asymptotic behavior.},
asymptotically for short distances, and at the leading order in the large-N limit, has the following spectral representation and asymptotic behavior in the coordinate representation, for $x\neq 0$ \footnote{For $x\neq 0$ no contact term (i.e. distribution supported at $x=0$) occurs, and there are no convergence problems for the spectral sum and the spectral integral in Equation \ref{eq:01} provided they are performed after the Fourier transform to the coordinate representation \cite{MBN}.}:
 \bea \label{eq:01}
&&\langle \mathcal{O}^{(s)}(x) \mathcal{O}^{(s)}(0) \rangle_{conn} \sim \sum_{n=1}^{\infty} \int  P^{(s)} \big(\frac{p_{\alpha}}{m^{(s)}_n}\big) \frac{m^{(s)2D-4}_n Z_n^{(s)2} \rho_s^{-1}(m^{(s)2}_n)}{p^2+m^{(s)2}_n  } \,e^{ip\cdot x}d^4p \nonumber \\
&& \sim  \int_{ m^{(s)2}_1}^{\infty}  \int P^{(s)} \big(\frac{p_{\alpha}}{p} \big)  \, p^{2D-4}   \frac{Z^{(s)2}(m)    }{p^2+m^{2}  } \,e^{ip\cdot x}d^4p \, dm^2  \nonumber \\
&&\sim 
\frac{\mathcal{P}^{(s)} \big(\frac{x_{\alpha}}{x}\big)}{x^{2D}} Z^{(s)2}(x, \mu) \mathcal{G}^{(s)}(g(x))
\sim \frac{\mathcal{P}^{(s)} \big(\frac{x_{\alpha}}{x}\big)}{x^{2D}}  (\frac{g^2(x)}{g^2(\mu)})^{\frac{\gamma_0}{\beta_0}} \nonumber \\
&&\sim \frac{\mathcal{P}^{(s)} \big(\frac{x_{\alpha}}{x}\big)}{x^{2D}} \Biggl(\frac{1}{\beta_0\log(\frac{1}{x^2 \Lambda^2_{QCD}})}\biggl(1-\frac{\beta_1}{\beta_0^2}\frac{\log\log(\frac{1}{x^2 \Lambda^2_{QCD}})}{\log(\frac{1}{x^2 \Lambda^2_{QCD}})}\biggr)\Biggr)^{\frac{\gamma_0}{\beta_0}}
\eea
where the infinite diverging sequence $ \{ m^{(s)}_n \}$ is supposed to be characterized by a smooth RG-invariant asymptotic spectral density (possibly dependent on $\mathcal{O}^{(s)}$) of the masses squared $\rho_s(m^{2})=\frac{dn}{dm^2}$ \cite{MBN}, for large masses and fixed spin, with dimension of the inverse of a mass squared. \par
$ P^{(s)} \big( \frac{p_{\alpha}}{m^{(s)}_n} \big)$ is a dimensionless polynomial in the four momentum $p_{\alpha}$, that projects on the free propagator of spin $s$ and mass $m^{(s)}_n$, and $\gamma_{\mathcal{O}^{(s)}}(g)= - \frac{\partial \log Z^{(s)}}{\partial \log \mu}=-\gamma_{0} g^2 + O(g^4)$, with $Z_n^{(s)}$ the associated renormalization factor computed at the momentum scale $p^2=m^{(s)2}_n$: $Z_n^{(s)}\equiv Z^{(s)}(m^{ (s)}_n)= \exp{\int_{g (\mu)}^{g (m^{(s)}_n )} \frac{\gamma_{\mathcal{O}^{(s)}} (g)} {\beta(g)}dg}$.
The renormalization factors are fixed asymptotically for large $n$ to be:
\begin{equation}\label{eqn:zk_as_behav}
Z_n^{(s)2}\sim 
\Biggl[\frac{1}{\beta_0\log \frac{ m^{ (s) 2}_n }{ \Lambda^2_{QCD} }} \biggl(1-\frac{\beta_1}{\beta_0^2}\frac{\log\log \frac{ m^{ (s) 2}_n }{ \Lambda^2_{QCD} }}{\log \frac{ m^{ (s) 2}_n }{ \Lambda^2_{QCD} }}    + O(\frac{1}{\log \frac{ m^{ (s) 2}_n }{ \Lambda^2_{QCD} } } ) \biggr)\Biggr]^{\frac{\gamma_0}{\beta_0}}
\end{equation}
$P^{(s)} \big(\frac{p_{\alpha}}{p} \big)$ is the projector obtained substituting $-p^2$  to $m_n^2$ in  $P^{(s)} \big(\frac{p_{\alpha}}{m_n} \big)$ \footnote{We use Veltman conventions for Euclidean and Minkowski propagators of spin $s$ \cite{MBN}.}. This substitution in Equation \ref{eq:01} is an identity up to contact terms \cite{MBN}, that do not contribute for $x \neq 0$. \par
The second line in Equation \ref{eq:01} occurs because asymptotically, under mild assumptions \cite{MBN}, we can substitute to the \emph{discrete} sum the \emph{continuous} integral weighted by the spectral density.
Thus the asymptotic spectral representation depends only on the anomalous dimension but not on the spectral density. This integral form of the Kallen-Lehmann representation holds asymptotically in the UV also in the Veneziano Theory and in the complete theory, since it does not assume a discrete spectrum. \par
$\mathcal{P}^{(s)} \big(\frac{x_{\alpha}}{x}\big)$ is the dimensionless spin projector in the coordinate representation in the conformal limit. The RG-invariant function of the running coupling only, $\mathcal{G}^{(s)}(g(x))$, admits the expansion: $\mathcal{G}^{(s)}(g(x))= const(1+ O(g^2(x)))$. \par
Indeed, perturbatively at the lowest non-trivial order the correlator of a hermitian operator in the coordinate representation must be exactly conformal and non-vanishing in a massless QCD-like theory, because the two-point correlator of a non-zero hermitian operator cannot vanish in a unitary conformal theory. \par
In fact, the coordinate representation is the most fundamental for deriving \cite{MBN} the asymptotic theorem, because only in the coordinate representation the operators are multiplicatively renormalizable, since for $x \neq 0$ no further additive renormalization due to possibly divergent contact terms may arise. \par
The asymptotic structure of the bare correlators in the complete theory follows from Equation \ref{eq:01} dividing by the asymptotic multiplicative renormalization factor of the complete theory $(\frac{g^2(\Lambda)}{g^2(\mu)})^{\frac{\gamma_0}{\beta_0}}$ :
$ 
\langle \mathcal{O}^{(s)}(x) \mathcal{O}^{(s)}(0) \rangle_{bare}  \sim \frac{\mathcal{P}^{(s)} \big(\frac{x_{\alpha}}{x}\big)}{x^{2D}} (\frac{g^2(x)}{g^2(\Lambda)})^{\frac{\gamma_0}{\beta_0}} 
$. \par
Reinserting the Planar multiplicative renormalization necessary to make finite the Planar correlator, we get in both 't Hooft and Veneziano Planar expansions (the superscript $\mathcal{P}$ stays for P or VP):
\bea \label{eq:03}
&&\langle \mathcal{O}^{(s)}(x) \mathcal{O}^{(s)}(0) \rangle_{conn}  \sim \frac{ \mathcal{P}^{(s)} \big(\frac{x_{\alpha}}{x}\big) }{x^{2D}} (\frac{g^2(\Lambda)}{g^2(\mu)})^{ \frac{\gamma^{\mathcal{P}}_0}   { \beta^{\mathcal{P}}_0 }}
(\frac{g^2(x)}{g^2(\Lambda)})^{\frac{\gamma_0}{\beta_0}} \nonumber \\
&&= \frac{\mathcal{P}^{(s)} \big(\frac{x_{\alpha}}{x}\big)}{x^{2D}}   (\frac{g^2(\Lambda)}{g^2(\mu)})^{ \frac{\gamma^{\mathcal{P}}_0}   { \beta^{\mathcal{P}}_0 } } 
(\frac{g^2(x)}{g^2(\Lambda)})^{      \frac{\gamma^{\mathcal{P}}_0}   { \beta^{\mathcal{P}}_0 }      }  (\frac{g^2(x)}{g^2(\Lambda)})^{\frac{\gamma_0}{\beta_0} -  \frac{\gamma^{\mathcal{P}}_0}   { \beta^{\mathcal{P}}_0 } }\nonumber \\
&& \sim \langle \mathcal{O}^{(s)}(x) \mathcal{O}^{(s)}(0) \rangle^{\mathcal{P}}\, (1 +  (\frac{\gamma_0}{\beta_0}- { \frac{\gamma^{\mathcal{P}}_0}   { \beta^{\mathcal{P}}_0 }} ) \log (\frac{g^2(x)}{g^2(\Lambda)})) \nonumber \\
&& \sim \langle \mathcal{O}^{(s)}(x) \mathcal{O}^{(s)}(0) \rangle^{\mathcal{P}} \, (1 +  (   \frac{\gamma_0}{\beta_0}- { \frac{\gamma^{\mathcal{P}}_0}   { \beta^{\mathcal{P}}_0 }}              ) \log (\frac{\log(\frac{\Lambda^2}{\Lambda^2_{QCD}})}{\log(\frac{1}{x^2 \Lambda^2_{QCD}})})) 
\eea
Thus the expansion of the correlators around the Planar Theory has in general loglog divergences due to the $\frac{1}{N}$ corrections to the anomalous dimensions. Remarkably, the correlator of $\Tr F^2$:
\bea \label{U}
&&\langle \Tr F^2(x) \Tr F^2(0) \rangle_{conn} \sim \frac{1}{x^{8}} (\frac{g^4(x)}{g^4(\mu)}) \nonumber \\
&&\sim \frac{1}{x^{8}} \Biggl(\frac{1}{\beta_0\log(\frac{1}{x^2 \Lambda^2_{QCD}})}\biggl(1-\frac{\beta_1}{\beta_0^2}\frac{\log\log(\frac{1}{x^2 \Lambda^2_{QCD}})}{\log(\frac{1}{x^2 \Lambda^2_{QCD}})}\biggr)\Biggr)^{2} 
\eea
has no such loglog corrections in the 't Hooft and Veneziano expansions, since $\gamma_0=2\beta_0$ for $\Tr F^2$, both in the complete theory and in the Planar Theory, and thus the change of the anomalous dimension is always compensated by the change of the beta function. Hence the only renormalization in Equation \ref{U} is due to the $\frac{1}{N}$ expansion of $\Lambda_{QCD}$ described in Section \ref{2}.

\section{Large-N massless QCD counterterms from a low-energy theorem, as opposed to perturbation theory} \label{4}

A new version of a NSVZ low-energy theorem is obtained as follows. For a set of operators $\mathcal{O}_i$, deriving:
\begin{equation}\label{def}
\braket{ \mathcal{O}_1 \cdots  \mathcal{O}_i }=\frac{\int  \mathcal{O}_1 \cdots  \mathcal{O}_i e^{-\frac{N}{2g^2} \int \Tr F^2(x)d^4x+\cdots}}
{\int e^{-\frac{N}{2g^2}\int \Tr F^2(x)d^4x +\cdots}}
\end{equation}
with respect to $-\frac{1}{g^2}$, we get:
\begin{equation}
 \frac{\partial\braket{\mathcal{O}_1 \cdots  \mathcal{O}_i }}{\partial \log g}=
\frac{N}{g^2}\int \braket{ \mathcal{O}_1 \cdots  \mathcal{O}_i  \Tr F^2(x)} - \braket{ \mathcal{O}_1 \cdots  \mathcal{O}_i} \braket{\Tr F^2(x)}d^4x
\end{equation}
Since non-perturbatively in massless QCD-like theories the only parameter is $\Lambda_{QCD}$, we can trade $g$ for $\Lambda_{QCD}$ in the LHS:
$
\frac{\partial\braket{\mathcal{O}_1 \cdots  \mathcal{O}_i }}{\partial (-\frac{1}{g^2})}= \frac{\partial\braket{\mathcal{O}_1 \cdots  \mathcal{O}_i }}{\partial \Lambda_{QCD}} \frac{\partial  \Lambda_{QCD}}{\partial (-\frac{1}{g^2})}
$. Employing the defining relation: $(\frac{\partial}{\partial \log \Lambda}+\beta(g)\frac{\partial}{\partial g})\Lambda_{QCD}=0$, with $\beta(g)=-\beta_0 g^3-\beta_1g^5+\cdots$, we obtain:
$
\frac{\partial \Lambda_{QCD}}{\partial(-\frac{1}{g^2})}=\frac{g^3}{2}\frac{\partial\Lambda_{QCD}}{\partial g}=
-\frac{g^3}{2\beta(g)}\frac{\partial \Lambda_{QCD}}{\partial \log \Lambda}=
-\frac{g^3}{2\beta(g)}\Lambda_{QCD}
$,
where the last identity follows from the relation: $\Lambda_{QCD}=\Lambda f(g)= e^{\log \Lambda} f(g)$, for some function $f(g)$. Hence we get a NSVZ low-energy theorem:
\begin{equation} \label{low}
\frac{\partial\braket{\mathcal{O}_1 \cdots  \mathcal{O}_i }}{\partial \log \Lambda_{QCD}} =
-\frac{N\beta(g)}{g^3}\int \braket{ \mathcal{O}_1 \cdots  \mathcal{O}_i  \Tr F^2(x)} - \braket{ \mathcal{O}_1 \cdots  \mathcal{O}_i} \braket{\Tr F^2(x)}d^4x
\end{equation}
Now we specialize to multiplicatively renormalized operators in the Planar Theory, $\mathcal{O}_i=\Tr F^2$, in such a way that the only source of divergences is the renormalization of $\Lambda_{QCD}$ (see the comment below Equation \ref{U}), being the combination $\frac{N\beta(g)}{g^3} \Tr F^2(x)$ already RG invariant \footnote{The unusual power of $g$ in front of $\Tr F^2(x)$ is due to the non-canonical normalization of the action in Equation \ref{def}.}. 
Therefore, the divergent part of the correlator at the lowest $\frac{1}{N}$ order is:
$\big[\braket{\mathcal{O}_1 \cdots  \mathcal{O}_i }^{\mathcal{NP}}\big]_{div} =
\frac{\partial\braket{\mathcal{O}_1 \cdots  \mathcal{O}_i }^{\mathcal{P}}}{\partial  \Lambda_{QCD}} \Lambda_{QCD}^{\mathcal{NP}}
$,
where $\mathcal{P}=P,VP$, $\Lambda_{QCD}^{NP}= \frac{\beta_0^{NP}}{\beta_0^{P}} \Lambda^P_{QCD} \log(\frac{\Lambda}{ \Lambda^P_{QCD}}) +\cdots $, and
$\Lambda_{QCD}^{NVP}=  \frac{\beta^{NVP}_1}{ 2\beta_0^2}\Lambda^{VP}_{QCD} \log\log(\frac{\Lambda}{ \Lambda^{VP}_{QCD}})+ \cdots$, in  't Hooft and Veneziano expansions of massless QCD respectively. \par
It follows from Equation \ref{low} that the divergent part of the correlator at leading order in the non-Planar Theory satisfies the new  low-energy theorem at large-N:
\bea
&&\big[\braket{\Tr F^2 \cdots \Tr F^2 }^{\mathcal{NP}}\big]_{div} \nonumber \\
&&=\frac{N\beta^{\mathcal{P}}(g) \Lambda_{QCD}^{\mathcal{NP}}}{g^3\Lambda_{QCD}^{\mathcal{P}}} \int
\braket{ \Tr F^2 \cdots  \Tr F^2}^{\mathcal{P}} \braket{\Tr F^2(x)}^{\mathcal{P}}-\braket{ \Tr F^2 \cdots  \Tr F^2  \Tr F^2(x)}^{\mathcal{P}} d^4x \nonumber \\
\eea
and thus arises, up to finite scheme-dependent corrections, from the divergent counterterm in the action: $- \frac{\beta^{\mathcal{P}}_0 N \Lambda_{QCD}^{\mathcal{NP}}}{\Lambda_{QCD}^{\mathcal{P}}} \int \Tr F^2(x)$.
In 't Hooft expansion: $- \frac{\beta^{\mathcal{P}}_0 N \Lambda_{QCD}^{\mathcal{NP}}}{\Lambda_{QCD}^{\mathcal{P}}} = - N \beta_0^{NP} [ \log(\frac{\Lambda}{ \Lambda^P_{QCD}})+ 
\frac{1}{2 \beta_0^{P}}(\beta_1^{NP}-\beta_0^{NP} \frac{\beta_1^{P}}{\beta_0^{P}}) \log  \log(\frac{\Lambda}{ \Lambda^P_{QCD}})]=  \frac{1}{(4 \pi)^2} \frac{2}{3} N_f \log(\frac{\Lambda}{ \Lambda^P_{QCD}})+\cdots$, that coincides exactly within the leading-log accuracy, perhaps as expected, with the perturbative counterterm arising from quark loops. The $ \log\log(\frac{\Lambda}{ \Lambda^P_{QCD}})$ counterterm follows from Equation \ref{alpha1} including the contributions from  $\beta_1$ by a straightforward but tedious computation.

\section{S-matrix in large-N massive QCD-like theories} \label{5}

We may wonder as to whether the results for massless theories described in Section \ref{2}, \ref{3}, \ref{4}, apply also to confining massive QCD-like theories, in particular to the large-N limit of massive QCD  \footnote{This point was raised by an anonymous referee.}. Introducing further mass scales is an additional complication, that may involve extra renormalizations associated to the mass parameters. However, the question that we answer in this Section is as to whether, supposing the further parameters have been already renormalized, the large-N expansion of the massive theory may get milder ultraviolet divergences than the massless one. \par
The simple answer is negative, provided a renormalization scheme exists in which the beta function is independent on the masses, as it is appropriate for the UV-complete massive theory, as opposed to the "low-energy" effective theory at scales much smaller than the masses: In such a scheme the renormalization of $\Lambda_{QCD}$ goes through exactly as in the massless theory, as described in Sections \ref{2}, \ref{3}, \ref{4}. An example is the $\overline{MS}$ scheme in massive QCD-like theories. \par
In particular, the large-N massive QCD S-matrix is renormalizable but not UV finite, as it is not its massless limit. Moreover, both the 't Hooft and the Veneziano expansions of the $\mathcal{N}=1$ SUSY massive QCD S-matrix in the Confining/Higgs phase \cite{Seiberg} are renormalizable but not UV finite, because the first-two coefficients of the beta function: $\beta_0 = \frac{1}{(4 \pi)^2} 3 - \frac{1}{(4 \pi)^2} \frac{N_f}{N}$, $\beta_1= \frac{1}{(4 \pi)^4} 6 - \frac{1}{(4 \pi)^4} (4 - \frac{2}{N^2}) \frac{N_f}{N}$, imply that $\beta_0^{NP}=- \frac{1}{(4 \pi)^2} \frac{N_f}{N} $ and $\beta_1^{NVP}=  \frac{2}{(4 \pi)^4} \frac{N_f}{N^3}$. \par
Another question is what happens regularizing and renormalizing a QCD-like theory by the embedding into an ultraviolet finite theory \footnote{This point was raised by the same anonymous referee.}, that for example is feasible concretely for $\mathcal{N}=1$ SUSY QCD with $1 \leq N_f \leq N$ and for $\mathcal{N}=1$ SUSY YM, by the embedding into a suitable finite $\mathcal{N}=2$ SUSY theory \cite{AM} containing massive multiplets on the order of $M$ that act as regulators, and may eventually be decoupled in the limit $M \rightarrow \infty$, in order to recover the original theory \cite{AM}. \par
In this respect the Veneziano limit of massive $\mathcal{N}=1$ SUSY QCD with $1 \leq N_f \leq N$ is particularly interesting, since in this case both the Veneziano Planar Theory and the next orders in the large-N expansion of the regularizing $\mathcal{N}=2$ theory are UV finite, since the beta function vanishes and the $\mathcal{N}=2$ SUSY is only softly broken by the massive multiplets \cite{AM}, the absence of divergences depending only on the vanishing of $\beta_0$ because of the $\mathcal{N}=2$ SUSY \cite{AM}. \par
However, being asymptotically conformal in the deep ultraviolet, the regularizing $\mathcal{N}=2$ theory is not $\mathcal{N}=1$ SUSY QCD, that instead is AF, that means that the conformal behavior is corrected in general in the correlators by fractional powers of logs, according to Equation \ref{eq:01}. \par
Thus, despite the finiteness of the $\mathcal{N}=2$ theory, what we want really to discover is the gauge-coupling renormalization of its $\mathcal{N}=1$ "low-energy limit" in the Veneziano expansion as the mass $M$ of the regulator multiplets goes to infinity. This is again the original question that we already answered above, the only difference being that the effective cutoff of the regularized $\mathcal{N}=1$ theory is now on the order of $M$ instead of $\Lambda$. 
Hence, though the regularized massive $\mathcal{N}=1$ SUSY QCD theory is finite for finite $M$, it is UV divergent in the Veneziano expansion as $M \rightarrow \infty$.\par
Finally, we should add that the asymptotic estimates for the correlators in the massless theory in Section \ref{3} apply without modification to massive QCD-like theories provided the massless limit exists smoothly, since in this case the leading UV asymptotics of the correlators is independent on the masses. Yet some modification may possibly arise in massive $\mathcal{N}=1$ SUSY QCD with $1 \leq N_f \leq N$, because the massless limit in the correlators may not be necessarily smooth, being the massles limit for certain SUSY meson one-point correlators divergent \cite{Seiberg}. 
\section{Acknowledgments}
We would like to thank Gabriele Veneziano for helpful comments.


\begin{thebibliography}{99}
\bibitem{H1} G. 't Hooft, \emph{A planar diagram theory for strong interactions}, Nucl. Phys. B {\bf 72} (1974) 461.
\bibitem{Veneziano0} G. Veneziano, \emph{Some Aspects of a Unified Approach to Gauge, Dual and Gribov Theories}, Nucl. Phys. B {\bf 117} (1976) 519.
\bibitem{AFB} M. Bochicchio, \emph{An asymptotic solution of large-$N$ $QCD$}, EPJ Web of Conferences {\bf 80}, (2014) 00010, \href{http://arxiv.org/abs/arxiv:1409.5144}{arXiv:1409.5144 [hep-th]}.
\bibitem{AF} M. Bochicchio, \emph{Asymptotic Freedom versus Open/Closed Duality in Large-N QCD}, \href{https://arxiv.org/abs/arXiv:1606.04546}{arXiv:1606.04546 [hep-th]}.
\bibitem{MBN} M. Bochicchio, \emph{Glueball and meson propagators of any spin in large-$N$ $QCD$}, Nucl. Phys. B {\bf 875} (2013) 621, \href{https://arxiv.org/abs/1305.0273}{arXiv:1305.0273 [hep-th]}.
\bibitem{MBM} M. Bochicchio, S. P. Muscinelli, \emph{Ultraviolet asymptotics of glueball propagators}, JHEP {\bf 08} (2013) 064, \href{https://arxiv.org/abs/1304.6409}{arXiv:1304.6409 [hep-th]}. 
\bibitem{Seiberg} K. Intriligator, N. Seiberg, \emph{Lectures on supersymmetric gauge theories and electric-magnetic duality}, Nucl. Phys. Proc. Suppl. B {\bf 45} (1996) 1, \href{https://arxiv.org/abs/arXiv:hep-th/9509066}{arXiv:hep-th/9509066 [hep-th]}.
\bibitem{AM} N. Arkani-Hamed, H. Murayama, \emph{Holomorphy, Rescaling Anomalies and Exact beta Functions in Supersymmetric Gauge Theories}, JHEP {\bf 06} (2000) 030, \href{https://arxiv.org/abs/arXiv:hep-th/9707133}{arXiv:9707133 [hep-th]}.
\end{thebibliography}
\end{document}